\documentclass[aps,prc,superscriptaddress,preprint,nofootinbib]{revtex4}
\usepackage{epsfig}
\usepackage{subfigure}
\usepackage{amsmath,amssymb}

\usepackage{changes}

\definechangesauthor[name={Yeunhwan}, color=magenta]{YH}

\begin{document}
\title{Tidal Deformability of Neutron Stars with Realistic Nuclear Energy Density Functionals}
%\deleted[id=YH]{most} realistic nuclear energy density functionals}
%
\author{Young-Min Kim}
\affiliation{School of Natural Science,
Ulsan National Institute of Science and Technology (UNIST),
Ulsan 44919, Korea}
\author{Yeunhwan Lim}
\affiliation{Cyclotron Institute, Texas A\&M University, College Station, TX 77843, USA}
\author{Kyujin Kwak}
\affiliation{School of Natural Science,
Ulsan National Institute of Science and Technology (UNIST),
Ulsan 44919, Korea}
\author{Chang Ho Hyun}
\affiliation{Department of Physics Education,
Daegu University, Gyeongsan 38453, Korea}
\author{Chang-Hwan Lee}
\affiliation{Department of Physics,
Pusan National University, Busan 46241, Korea}

\begin{abstract}
We investigate the constraints on the mass and radius of neutron stars by considering
the tidal deformability in the merge of neutron star binaries.
In order to extract the most reliable range of uncertainty from theory,
we employ models based upon the Skyrme force and density functional theory and  
select models that are consistent with empirical data of finite nuclei,
measured properties of nuclear matter around the saturation density,
and observation of the maximum mass of neutron stars.
From the selected models, we calculate the Love number $k_2$, dimensionless tidal deformability $\Lambda$,
and mass-weighted deformability $\tilde{\Lambda}$ in the binary system.
We find that all the models considered in this work
give $\tilde{\Lambda}$ less than 800 which is the upper limit
obtained from the measurement of GW170817.
However, the model dependence of tidal deformability 
is manifest such that our results on the tidal deformability exhibit critical 
sensitivity to the size of neutron stars.
\end{abstract}

\date{\today}

\maketitle

\section{Introduction}\label{sec:intro}

A series of detections of gravitational wave (GW170817), short $\gamma$-ray burst (GRB 170817A), and electromagnetic
waves from X-ray to radio bands (AT2017gfo) made it possible to identify the source of the event as the
coalescence of two neutron stars \cite{170817,170817GRB,170817AT}.
From the detection of the phase evolution of gravitational wave during the inspiral phase of merge,
one can extract tidal deformability, which characterizes the quadrupole deformation of a neutron star
in response to the external quadrupolar gravitational field (i.e., order $l=2$) \cite{flanagan2008, hinderer2008}.
The extent of deformation is closely related to the structure of a neutron star and 
equation of state (EoS) of matter that consists of the neutron star.
Measurement of GW170817 at advanced LIGO and Virgo and its analysis put strong limit on the 
range of tidal deformability such as $\tilde{\Lambda} \leq 800$ and $\Lambda(1.4 M_\odot)  \leq 800$ 
for the case of low-spin priors, where $\tilde{\Lambda}$ is  combined dimensionless tidal deformability of the binary system and  $\Lambda(1.4 M_\odot)$ is dimensionless tidal deformability of a single neutron star having a mass of $1.4 M_\odot$~\cite{170817}.
The upper limits on $\tilde{\Lambda}$  and $\Lambda(1.4 M_\odot)$ obtained from GW170817 exclude 
stiff EoSs which predict their values larger than 800. 
In contrast, observations of massive neutron stars with mass $(1.97 \pm 0.04) M_\odot$ \cite{a} and 
$(2.01 \pm 0.04) M_\odot$ \cite{b} prefer stiff EoSs, 
ruling out soft EoSs which cannot produce maximum mass larger than these values.
Compared with the mass measurement alone, measuring $\tilde{\Lambda}$ has a more advantageous implication that it provides simultaneous constraint on both mass and radius.
For example, Ref.~\cite{at2017} performs a model analysis of 
UV/optical/IR counterpart of GW170817 and obtains a lower bound $\tilde{\Lambda} \geq 400$.
This lower limit rules out models that predict neutron star with too small size. 
With both upper and lower bounds of $\tilde{\Lambda}$, it is now possible to put unprecedented 
strong constraint on the EoS of dense nuclear matter.

EoS of nuclear matter is developed and tested in various ways, but 
data of finite
nuclei provide the most essential and fundamental laboratory for them. 
Depending on the choice of input data and methods of fitting, the number of
models that are adjusted to the data of finite nuclei amounts to several hundreds~\cite{dutra}.
It has been known that 
most models are good at producing results consistent with the properties of stable nuclei
in the nuclear chart.
However, if one moves out of the valley of stability
(i.e., moves into the regime of unstable nuclei)
and/or calculates EoS at densities 
away from the saturation density ($\rho_0 = 0.16\, {\rm fm}^{-3}$), model predictions do not converge. 
In this case, models are tested against the properties of nuclear matter which are generally 
different from those of stable nuclei but can still be measured and/or derived from experiments. 
A good example can be found in Ref.~\cite{dutra} which 
calculates the properties of nuclear matter around the saturation density with
240 Skyrme force models and compares them with 11 experimental and/or empirical (i.e, derived from experiments) data (denoted as Exp/Emp).
It is found that only 16 models satisfy all 11 Exp/Emp constraints.
Additional test for EoS comes from the observation of neutron stars' mass. As mentioned earlier, 
imposing maximum mass of a neutron star equal to or larger than $2 M_\odot$ can rule out models that
cannot predict this value of maximum mass.
In summary, three sets of constraints, data of stable nuclei, properties of nuclear matter, 
and the maximum mass of
neutron stars, can be used simultaneously for the selection of models in 
a wide range of density and neutron-proton asymmetry because they provide stringent constraints 
together. 

In this work, we select some models from a group of nuclear energy density functional (EDF) models 
by applying the three constraints mentioned above and investigate the range of tidal deformability allowed by these selected (i.e., most realistic) nuclear EDF models.
Our selected models are based on the Skyrme force model and density functional theory (DFT). 
We select five models: 
(1) three models, SkI4 \cite{ski4}, SGI \cite{sg1}, and SLy4 \cite{sly4} from the Skyrme force models (2) one model, GSkI \cite{gsk1} from the generalized Skyrme force models (3) the KIDS model \cite{prc2018} from the DFT models. 
Note that KIDS incorporates a
perturbation scheme for the expansion of EDF. 
More details of these models are discussed in a later section. From these models, 
we obtain the combined dimensionless tidal deformability $\tilde{\Lambda}$
in the range of $360 \leq \tilde{\Lambda} ~(M_{\rm chirp}=1.188M_{\odot}) \leq 700$.
This result is consistent with that  from GW170817 and AT2017gfo, $400 \leq \tilde{\Lambda} \leq 800$.

We organize the paper as follows.
In Section II, we present definitions of basic quantities and equations relevant to 
tidal deformability.
In Section III, properties and behaviors of selected models are explained.
In Section IV, we present the results and discuss them in comparison with
measurements and predictions from other theoretical works.
Section V summarizes our work. 
% and suggests prospects for the future work.  

%--------------------------------
% Formalism section
%--------------------------------

\section{Formalism}

%$G=c=1$ unit.

In this section, we briefly summarize the formalism to obtain the tidal deformability of neutron stars.
The tidal deformability ($\lambda$)
parameterizes the connection between the external quadrupolar tidal field (${\cal E}_{ij}$) and the induced quadrupole moment of a neutron star ($Q_{ij}$) as
\begin{eqnarray}
Q_{ij} = - \lambda {\cal E}_{ij},
\end{eqnarray}
where  
\begin{eqnarray}
Q_{ij} &=& \int d^3x \delta\rho(x) \left(x_i x_j - \frac 13 r^2 \delta_{ij}\right), \nonumber\\
{\cal E}_{ij} &=& \frac{\partial^2\Phi_{\rm ext}}{\partial x^i \partial x^j},
\end{eqnarray}
and $\Phi_{\rm ext}$ is an external gravitational potential~\cite{flanagan2008,hinderer2008}. Note that $G=c=1$ units are used in this work and the typical value of $\lambda$ is ${\cal O} ( 10^{36}\ {\rm g\ cm^2\ s^2})$.
The tidal deformability can be represented by the dimensionless tidal Love number $k_2$,
\begin{eqnarray}
k_2 = \frac{3}{2}  \lambda R^{-5},
\end{eqnarray}
where $R$ is the radius of a neutron star.

The tidal Love number $k_2$ can be obtained by calculating
\begin{eqnarray}
k_2 &=& \frac{8 C^5}{5} (1-2 C)^2 \left[ 2 + 2C (y-1) -y \right]  {\cal Z}^{-1} ,
\end{eqnarray}
where $C=M/R$ is the compactness parameter of a neutron star with mass $M$,  and
\begin{eqnarray}
{\cal Z} &=& 2 C \left[ 6 -3 y + 3C (5y-8) \right] 
 + 4 C^3 \left[ 13 - 11y + C (3y-2) + 2 C^2 ( 1+y) \right] \nonumber\\
&& + 3 (1-2C^2) \left[2-y + 2C (y-1) \right] \ln (1-2C)
\end{eqnarray}
with $y=(r H^\prime/H) |_{r=R}$. Here the prime denotes derivative with respect to $r$ and
$H(r)$ is the solution of  2nd order differential equation obtained by taking a linear $l=2$ perturbation in the metric equation with Regge-Wheeler gauge~\cite{Hin10},
\begin{eqnarray}
%\frac{dH}{dr} &=& \beta \nonumber \\
%\frac{d\beta}{dr} & = & 
H^{\prime\prime} + H^\prime \left[\frac{2}{r} +  {\cal Q}_1 \right] + H \left[ {\cal Q}_0 - 4  \left(\nu^\prime\right)^2 \right] =0,
\label{eqH}
\end{eqnarray}
with
\begin{eqnarray}
{\cal Q}_1 &=& \left(1-\frac{2m}{r}\right)^{-1}  \left[ \frac{2m}{r^2} - 4\pi  r(\epsilon - p) \right], \nonumber\\
{\cal Q}_0 &=& \left(1-\frac{2m}{r}\right)^{-1}  \left[ 4\pi ( 5\epsilon  + 9 p) + 4\pi \frac{d\epsilon}{dp} (\epsilon + p) - \frac{6}{r^2}  \right],\nonumber\\
\nu^\prime &=& \left(1-\frac{2m}{r}\right)^{-1} \left( \frac{m}{r^2} + 4\pi r p \right).
\end{eqnarray}
For a given EoS, we can obtain the enclosed mass $m(r)$, energy density $\epsilon(r)$, and pressure $p(r)$ inside a neutron star by solving the following TOV equations, 
%{Here the enclosed mass $m(r)$, energy density $\epsilon(r)$ and pressure $p(r)$ inside neutron %stars can be obtained} by solving the Tolman-Oppenheimer-Volkoff (TOV) equations,
\begin{eqnarray}
\frac{dp}{dr} &=& - \left(1-\frac{2m}{r}\right)^{-1}  (\epsilon + p) \left(\frac{m}{r^2} + 4\pi r p \right), \nonumber\\
\frac{dm}{dr} &=& 4\pi r^2 \epsilon.
\label{eqtov}
\end{eqnarray}
Note that the tidal Love number and tidal deformability can be obtained by solving Eq.~(\ref{eqH}) and Eq.~(\ref{eqtov}) simultaneously. 
%{Hence, once a neutron star EoS is obtained by solving TOV equations, the tidal Love number and tidal deformability can be easily calculated.}

In the detection of gravitational waves, the tidal deformability is measured by accumulating the phase differences of gravitational waves. Hence, the dimensionless deformability $\Lambda$ which naturally appears in the dimensionless phase is commonly used~\cite{Fav14,170817};
\begin{eqnarray}
\Lambda = \frac{2}{3} \left(\frac{R}{M}\right)^5 k_2 \approx 9495 \times \left(\frac{R_{10\rm km}}{M_{M_\odot}}\right)^5 k_2,
\label{eqL}
\end{eqnarray}
where $R_{\rm 10 km}$ and $M_{M_\odot}$ are radius and mass of a neutron star in units of 10 km and 
solar mass ($M_\odot$), respectively.
Note that the large coeffienent in Eq.~(\ref{eqL}) makes the deformability measurement possible. 
In a binary system, mass weighted tidal deformability $\tilde \Lambda$ is defined with individual deformabilities, $\Lambda_1$ and $\Lambda_2$, as
\begin{eqnarray}
\tilde\Lambda = \frac{16}{13} \frac{(M_1+12 M_2) M_1^4 \Lambda_1 + (M_2+12 M_1) M_2^4 \Lambda_2}{(M_1+M_2)^5},
\label{eqLtilde}
\end{eqnarray} 
where $M_1$ and $M_2$ are masses of two neutron stars in the binary~\cite{170817}.
Note that $\tilde\Lambda = \Lambda_1 = \Lambda_2$ when $M_1 = M_2$.

%----------------------------------
% end fo formalism
%----------------------------------

\section{Model}\label{sec:model}
\begin{table}[t]
\begin{center}
\begin{tabular}{c|c|c|c|c|c|c|c|c}\hline\hline
Model &$\rho_0$ & $E_0$  & $K_0$ & $-Q_0$ & $J$ & $L$ & $-K_\tau$ & $M_{\rm max}$ \\ \hline
Exp/Emp & $\simeq$ 0.16 & $\simeq$ 16.0 & $200\sim260$ & $200\sim1200$ & $30\sim35$ & $40\sim76$ & $372\sim760$ & $\geq 1.93 $ \\ %\sim2.05$ \\
CSkP & - & - & $202.0 \sim 240.3$ & $362.5 \sim 425.6$ & $30.0 \sim 35.5$ & $48.6 \sim 67.1$ &
$360.1 \sim 407.1$ & - 
\\
GSkI & 0.159 & 16.02 & 230.2 & 405.6 & 32.0 & 63.5 & 364.2 & 1.98 \\ 
SLy4 & 0.160 & 15.97 & 229.9 & 363.1 & 32.0 & 45.9 & 322.8 & 2.07 \\
SkI4 & 0.160 & 15.95 & 248.0 & 331.2 & 29.5 & 60.4 & 322.2 & 2.19 \\
SGI & 0.154 & 15.89 & 261.8 & 297.9 & 28.3 & 63.9 & 362.5  & 2.25 \\
KIDS & 0.160 & 16.00 & 240.0 & 372.7 & 32.8 & 49.1 & 375.1 & 2.14 \\ \hline\hline
\end{tabular}
\end{center}
\caption{Properties of nuclear matter and maximum mass of a neutron star 
calculated with five selected models. Saturation density ($\rho_0$) is in unit of fm$^{-3}$. 
Exp/Emp and CSkP values are quoted from Ref.~\cite{dutra}.
$E_0$, $K_0$, and $Q_0$ are binding energy per particle, compression modulus, and skewness 
(the third derivative of energy per particle) at the saturation density in the symmetric nuclear matter, 
respectively.  
%{\bf (Is this sentence a correct statement?)}. 
$J$, $L$, and $K_\tau$ are related to the symmetry energy of nuclear matter (see the text for details). 
$E_0$, $K_0$, $Q_0$, $J$, $L$, and $K_\tau$ are in unit of MeV and 
the maximum mass of neutron star ($M_{\rm max}$) in unit of the solar mass ($M_\odot$).}
\label{table:model}
\end{table}

As mentioned in Sec.~\ref{sec:intro}, we consider four models, SLy4, SkI4, SGI, and GSkI 
based upon the Skyrme force model, and a recently developed generalized EDF model, KIDS 
in this work.
We applied three rules to select these models: 
(i) data of finite nuclei, 
(ii) properties of nuclear matter near saturation, and 
(iii) the maximum mass of neutron stars.

The Skyrme force models have an advantage over the KIDS model that 
they automatically satisfy well-known properties of finite nuclei
since their model parameters are designed to fit them. 
In constrast, the KIDS model assumes rules to expand nuclear EDF 
in powers of the Fermi momentum $k_F$
in homogeneous nuclear matter and determines the 
model parameters in order to reproduce well-known EoS of infinite nuclear matter.
After the model parameters are fixed in this way, the model is applied back to calculating 
the properties of finite nuclei.
As a result, in many cases, the properties of finite nuclei calculated with the KIDS model 
are pure prediction. 
Therefore, in principle they don't have to be as good as those obtained with Skyrme force models
in explaining the properties of finite nuclei.
%
%do not agree to the data of finite nuclei as well as those with the Skyrme force models do.
%
However, it is shown that the KIDS model produces a good hierarchical structure in powers of $k_F$ 
in the homogeneous nuclear matter \cite{prc2018} 
and reproduces the properties of magic nuclei to an accuracy as good as conventional models \cite{acta2017,sm2017}.

Table \ref{table:model} summarizes the properties of nuclear matter at the saturation density 
and the maximum mass of neutron star calculated from five selected models that have passed 
our three criteria. 
Exp/Emp values are also included in the table 
for comparison. These values were used for the model selection.  
At the saturation density, the baryon number density of 
nuclear matter reaches a value of $\rho_0 \approx 0.16~{\rm fm}^{-3}$. 
In the symmetric nuclear matter, binding energy (per particle), 
compression modulus (second derivative of energy 
per particle), and skewness (third derivative of energy per particle) are denoted as $E_0$, $K_0$, and $Q_0$, 
respectively, at the saturation density ($\rho_0$). 
Nuclear symmetry energy $S(\rho)$, 
which is caused by the difference in the number of neutron and proton in the nuclear matter, 
depends on the particle number density ($\rho$) and 
is conventionally expanded in powers of $x \equiv (\rho - \rho_0)/3 \rho_0$ as 
\begin{eqnarray}
S(\rho) = J + L\, x + \frac{1}{2} K_{\rm sym}\, x^2 + {\cal O}(x^3)~, 
\end{eqnarray}
where $J$, $L$, and $K_{\rm sym}$ are measured and/or derived from experiments on 
non-symmetric nuclear matter like (heavy) nuclei having different numbers of protons and neutrons. 
$K_\tau$ in Tab.~\ref{table:model} is defined as
\begin{equation}
K_\tau \equiv K_{\rm sym} - 6 L - \frac{Q_0}{K_0} L~.
\end{equation}
Table \ref{table:model} shows that the properties of nuclear matter calculated with 
five selected models are in good agreement with those of Exp/Emp.

Considering observational uncertainties in the maximum mass of neutron star, 
the lowest limit becomes $1.93 M_\odot$, which we use as a third criterion to 
select models. Note that five models in Table \ref{table:model} can produce the maximum 
mass above this lowest limit. 
As indicated in Sec.~\ref{sec:intro}, only 16 out of 240 Skryme force models 
satisfy all of 11 constraints (denoted by CSkP)
based upon the Exp/Emp data of nuclear matter 
\cite{dutra}.
However, even these 16 {\it ``successful"} models predict a wide range of the maximum mass of 
neutron star. 
For instance, the GSkI model satisfies the maximum mass constraint, while the LNS model \cite{lns2006},  
% {\bf (may need to add a reference for this model because we did for the other models)} 
which belongs to the 16 {\it ``successful"} models, gives the maximum mass less than $1.8 M_\odot$.
This difference is mainly caused by the fact that the behavior of EoS at high densities varies significantly model to model because 
it has been known that the maximum mass of neutron star is sensitive to EoS at high densities. 
We note that among the five models in the current work, 
GSkI and KIDS satisfy all of 11 constraints of nuclear matter in \cite{dutra}. 
(Note that they also satisfy the other two criteria, properties of finite nuclei and the 
maximum mass of neutron star.)
%{\bf Question: Seven items in Table 1
%are different from 11 constraints in the reference? If not, we had better say something else.}
%as well as the properties of nuclei and the neutron star maximum mass.
Thus, these two models could be regarded as the most refined model 
that is consistent with presently available
data of finite nuclei, nuclear matter, and neutron star.

It is worth emphasizing again that the maximum mass of neutron star 
provides an essential criterion for EoS because it allows us to select 
specific models whose behavior at the densities much higher than $\rho_0$ must 
result in the proper maximum mass of neutron star. 
As we will see from the results in the following section, tidal deformability can 
also provide
more stringent constraints on both mass and radius of a neutron star.

% \added{Comments (YH) : I don't think $Q_0$ and $K_\tau$ are constrained by experiments yet. 
% The only constraints for them are obtained from model calculations not directly from
% nuclear experiments. For example, $K_0$ can be limited by dipole polarizability. But
% not for $Q_0$ and $K_\tau$.
% Am I wrong?}

\section{Result}

\begin{figure}[h]
\subfigure[~Mass vs. Central Density]{\includegraphics[width=8.5cm]{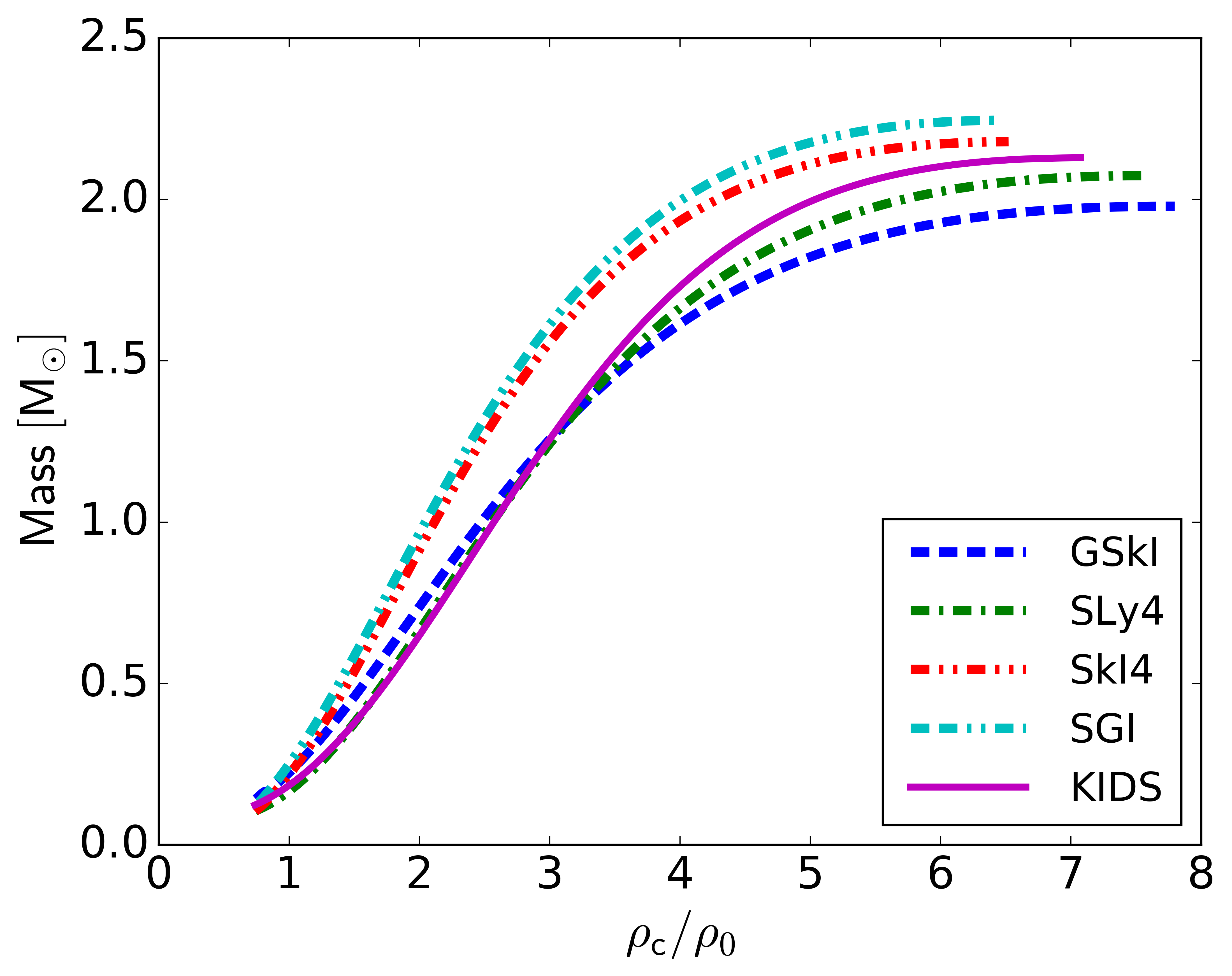}}
\subfigure[~Mass vs. Radius]{\includegraphics[width=8.5cm]{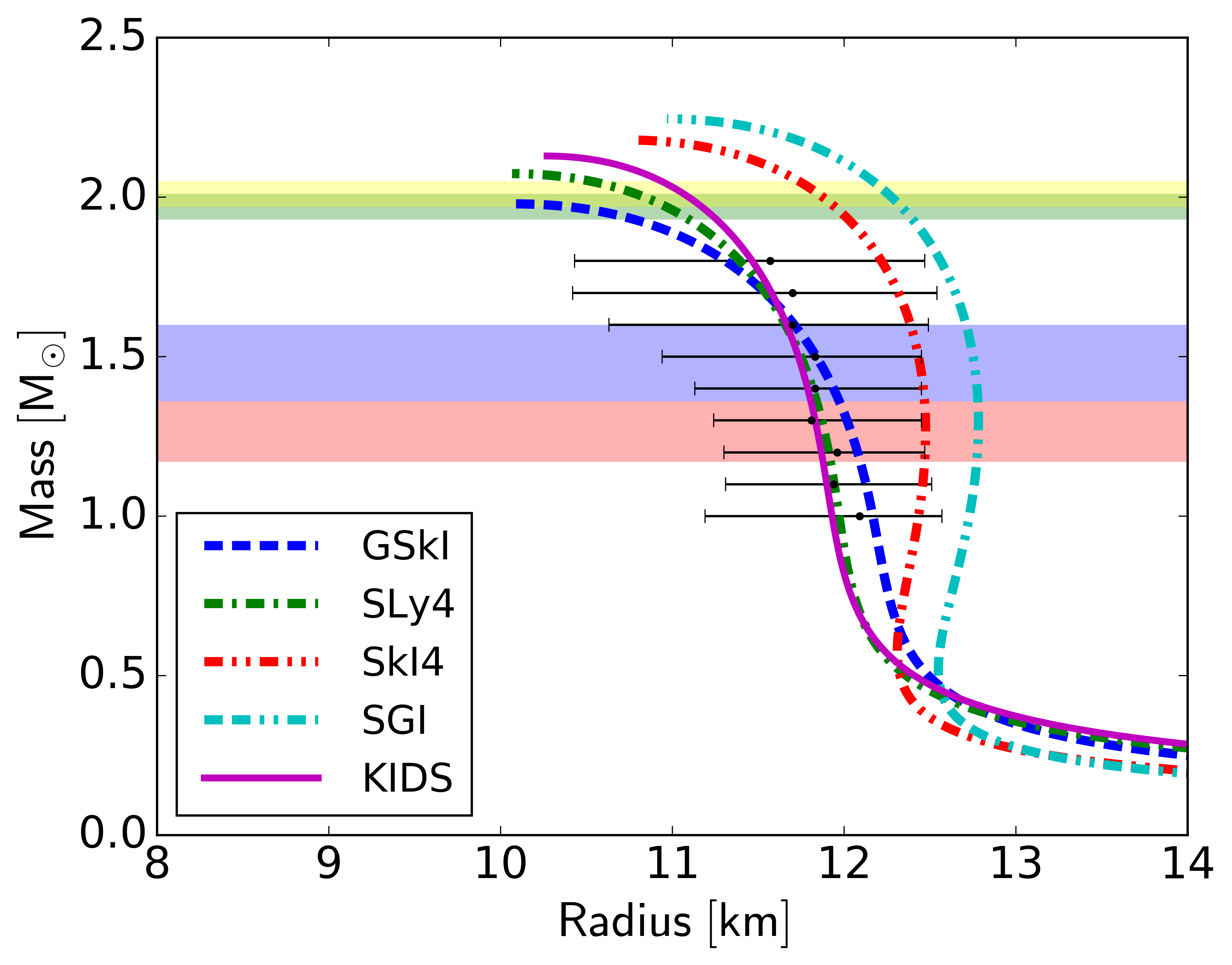}}
\label{fig1b}
\caption{Bulk properties of neutron stars 
predicted from the five selected EoS models described in Sec.~\ref{sec:model}. 
Top (a) and bottom (b) panels show the neutron star mass as a function of the 
central density and the mass-radius 
relationship, respectively. In the bottom panel, yellow and green bands represent the $2 M_{\odot}$ constraint \cite{a,b}, while blue and red bands indicate a neutron star mass of primary (M1) and secondary (M2) in the binary merger, respectively, estimated from GW170817 \cite{170817}. Horizontal error bars 
in the bottom panel (b) indicate the probable ($2 \sigma$ region) radii of neutron stars estimated from the 
Bayesian analyses of X-ray burst observations \cite{steiner2010}.}
\label{fig1}
\end{figure}

In order to obtain the results of tidal deformability, we solve Eq.~(\ref{eqH}) and 
Eq.~(\ref{eqtov}) simultaneously for five selected EoS models 
which satisfy the constraints of finite nuclei, nuclear matter, 
and the maximum mass of neutron star as described in the previous section. 
Before presenting the results of tidal deformability, 
we examine the bulk properties of neutron stars
predicted from the five selected EoS models in Fig.~\ref{fig1}. 
The upper panel, Fig.~\ref{fig1}(a), shows the neutron star mass $M$ 
as a function of the central density. 
As expected, all of five models reach the mass $2.0 M_{\odot}$, 
but the central density ($\rho_c$) for which the neutron star mass is equal to 
$2.0 M_{\odot}$ varies from model to model. 
Similarly, the central density at which the NS mass becomes $1.4 M_{\odot}$ 
also varies among the models. 
For example, $M = 1.4 M_{\odot}$ when $\rho_c \sim 2.7 \rho_0$ 
for SkI4 and SGI and $\rho_c \sim 3.3 \rho_0$ for GSkI, SLy4, and KIDS. 
The lower panel, Fig.~\ref{fig1}(b), plots the 
predicted mass and radius simultaneously. 
As in Fig.~\ref{fig1}(a), stiff (SkI4 and SGI) and soft (GSkI, SLy4, and KIDS) 
models behave differently for the mass-radius relation. 
Note that the three soft models satisfy the constraints of the mass-radius obtained from the Bayesian analyses of X-ray burst observations \cite{steiner2010}, better than the other two stiff models. 

%
%As shown in Fig.~\ref{fig1}(b), these EoSs satisfy a mass constraint, ${\rm M_{\rm NS,max}} \geq 2.0 M_{\odot}$ \cite{demorest2010,antoniadis2013} and radius predictions  are on 2$\sigma$ region estimated by a Bayesian analysis on X-ray burst observations \cite{steiner2010}.
%

\begin{figure}[h]
\includegraphics[width=8.5cm]{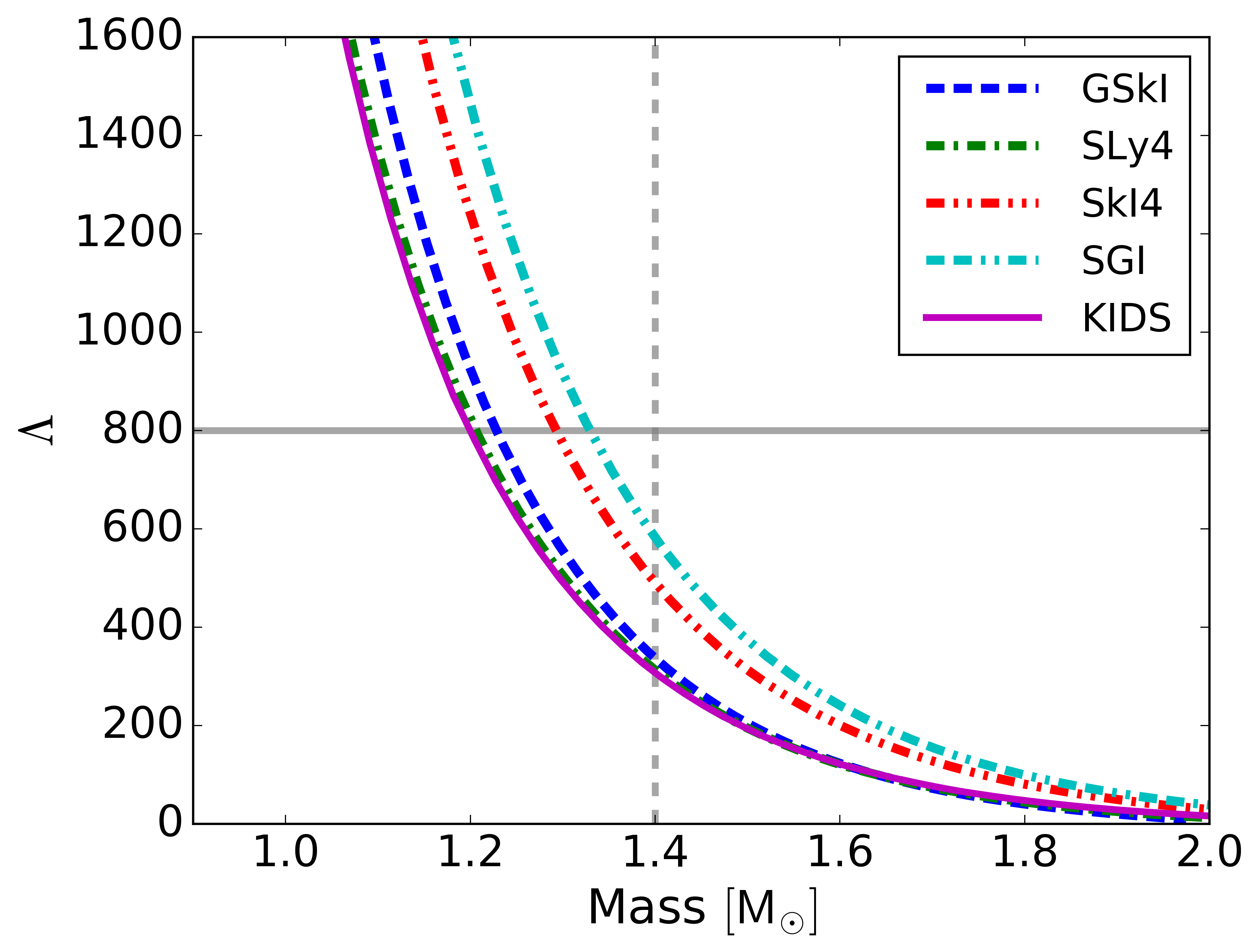}
\caption{Dimensionless tidal deformability ($\Lambda$) of a single neutron star as a function of mass. A horizontal grey line indicates the upper limit of $\Lambda$ estimated from GW170817 \cite{170817}. A vertical grey dotted line corresponds to 
1.4$M_{\odot}$.}
\label{fig2}
\end{figure}

The dimensionless tidal deformability ($\Lambda$) calculated for a single neutron star 
is shown as a function of the neutron star mass in Fig.~\ref{fig2}. 
For five models selected in this work, $\Lambda$ decreases as the neutron star mass increases. 
As in Fig.~\ref{fig1}, our five selected
models are divided into two groups, soft (GSkI, SLy4, and KIDS) and stiff (SkI4 and SGI). 
Note that for $M = 1.4 M_{\odot}$, the soft models predict $\Lambda < 400$ 
while the stiff models $\Lambda > 400$. 
However, all the five models satisfy the upper limit of the dimensionless tidal deformability 
($\Lambda < 800$) at $ M = 1.4 M_{\odot}$ which was estimated from GW170817 \cite{170817}.
For the quantitative comparison with other models, we provide the radius, tidal Love number, and 
dimensionful and dimensionless tidal deformability calculated for a single neutron star 
with $1.4 M_{\odot}$ from the five selected models in Table~\ref{table:radius}.

\begin{table}[t]
\begin{center}
\begin{tabular}{c|c|c|c|c|c}\hline\hline
 & GSkI & SLy4 & SkI4 & SGI & KIDS  \\ \hline
$R_{1.4M_\odot}$ [km] & 11.94 & 11.82 & 12.46 & 12.77 & 11.79  \\ 
$k_2(1.4M_\odot)$ &  0.079 & 0.077 & 0.092 & 0.097 & 0.076  \\ 
$\lambda(1.4M_\odot)$ [$\rm 10^{36} g~ cm^2 s^2$] & 1.906 & 1.770 & 2.772 & 3.292 & 1.737 \\
$\Lambda(1.4M_\odot)$ & 337.2 & 312.9 & 490.9 & 583.0 & 307.5 \\ \hline\hline
\end{tabular}
\end{center}
\caption{Radius, tidal Love number, and tidal deformability with and without dimension of a single neutron star with $M = 1.4 M_{\odot}$ calculated from our five EoS models.}
\label{table:radius}
\end{table}

\begin{figure}[h]
\subfigure[~Mass--weighted tidal deformability ($\tilde{\Lambda}$) as a function of mass ratio]{\includegraphics[width=8.5cm]{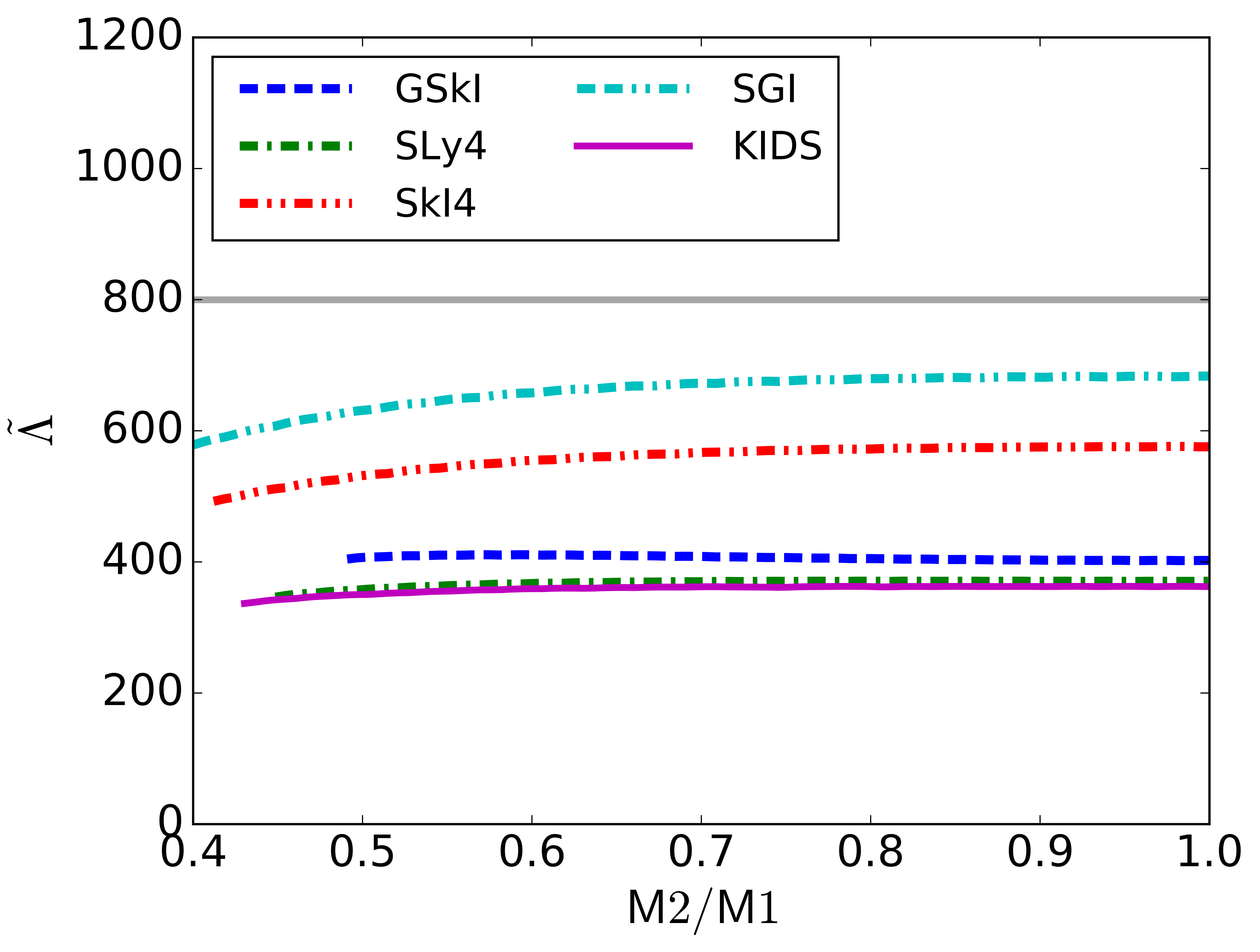}}
\subfigure[~Mass--weighted tidal deformability ($\tilde{\Lambda}$) in the $\Lambda_1$--$\Lambda_2$ space]{\includegraphics[width=8.5cm]{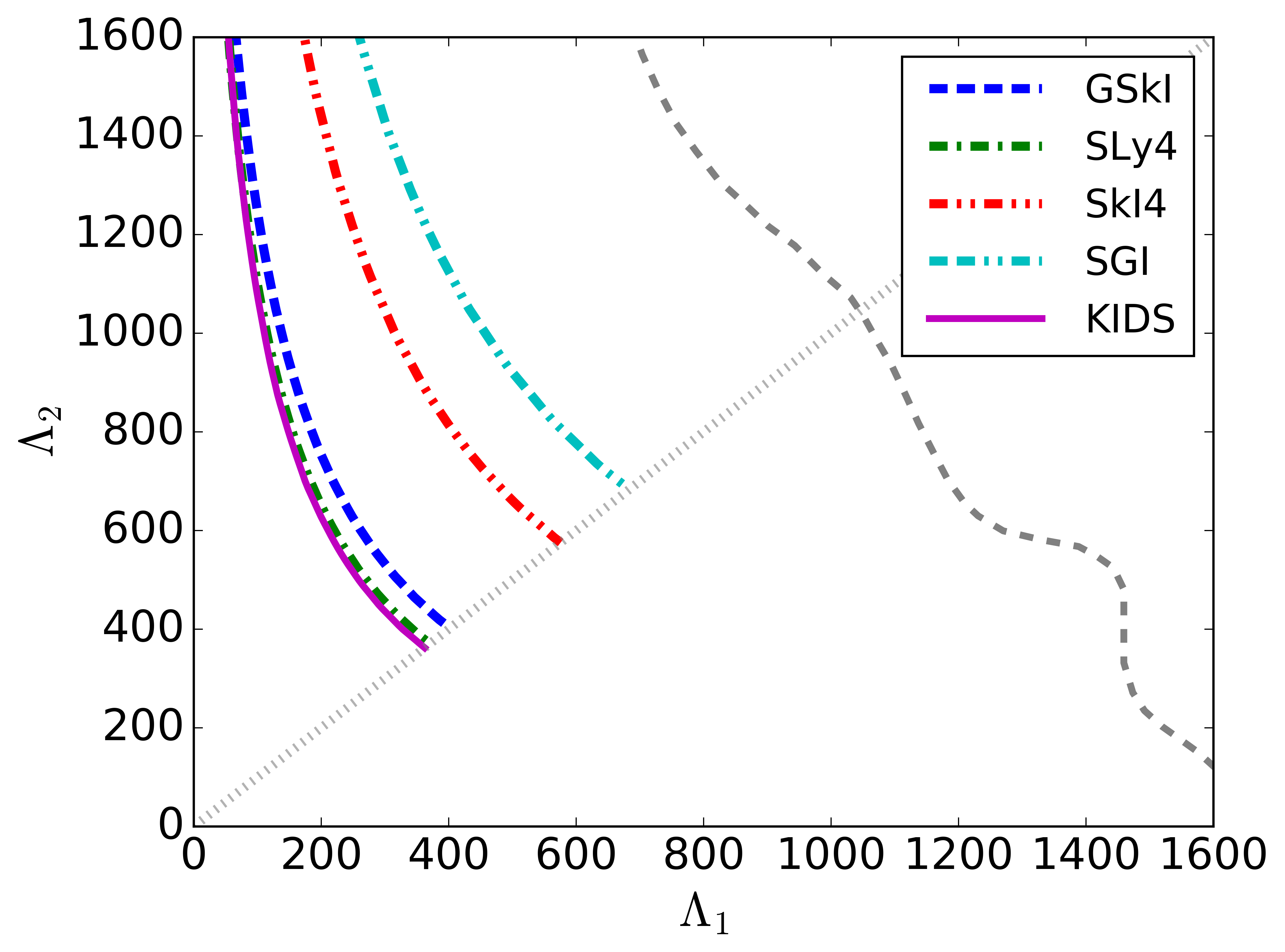}}
\caption{Mass-weighted tidal deformability ($\tilde{\Lambda}$) in the binary of two neutron stars. 
We fix the chirp mass of the binary at $M_{\rm chirp}=1.188 M_{\odot}$ as observed by LIGO/Virgo~\cite{170817} but allow the mass ratio to vary.
The top (a) and bottom (b) panels show $\tilde{\Lambda}$ as a function of 
mass ratio ($M_1/M_2$) and in the $\Lambda_1$--$\Lambda_2$ space, respectively. 
The horizontal grey line in the top panel is $\tilde{\Lambda}=800$, the upper limit estimated from GW170817. 
In the bottom panel, a similar upper limit is indicated by the grey dashed curve which represents the 90\% confidence limit of $\tilde{\Lambda} \leq 800$ in the $\Lambda_1$--$\Lambda_2$ space 
(i.e., the region  below the curve corresponds to $\tilde{\Lambda} \leq 800$ with 90\% confidence). 
The diagonal grey dotted line in the bottom panel corresponds to the case of an equal-mass binary, i.e., 
$M_1 = M_2 = 1.36 M_{\odot}$.
In this case, $\tilde{\Lambda}=\Lambda_1=\Lambda_2$.}
\label{fig4}
\end{figure}

The tidal effect measured in the LIGO/Virgo observations is obtained from the gravitational wave emitted from the merger of two neutron stars. Therefore the quantity most relevant to measurement is the mass-weighted tidal deformability ($\tilde{\Lambda}$) of a binary system.
The observations of GW170817 put an upper limit on $\tilde{\Lambda}$ such as $\tilde{\Lambda}  \leq 800$ for the low-spin prior. 
With the probable mass ratio $q = 0.7 \sim 1.0$ which is also estimated from GW170817 for the low-spin prior \cite{170817}, 
our selected five EoS models satisfy this upper limit of $\tilde{\Lambda}$ in the binary of two neutron stars (see the upper panel of Fig.~\ref{fig4}). 
Note that $\tilde{\Lambda}$ is calculated with Eq.~(\ref{eqLtilde}) for the merger of two neutron stars with masses $M_1$ and $M_2$. Similar to the behaviors of a single neutron star as seen in Fig.~\ref{fig2}, the two stiff models (SGI and SkI4) predict larger values of $\tilde{\Lambda}$ than the other three soft models 
(GSkI, SLy4, and KIDS). Note that the values of $\tilde{\Lambda}$ predicted from the soft models are marginal in comparison with the lower limit of $\tilde{\Lambda}$ which was estimated to be $\tilde{\Lambda} \geq 400$ in \cite{at2017}. The mass--weighted tidal deformability ($\tilde{\Lambda}$) can be also drawn as a contour in the $\Lambda_1$--$\Lambda_2$ space, where $\Lambda_1$ and $\Lambda_2$ are dimensionless tidal deformability of each neutron star in the binary. The lower panel of Fig.~\ref{fig4} shows the contour plots obtained from our five models which confirm the results seen in 
Fig.~\ref{fig4}(a).


\begin{figure}[h]
\includegraphics[width=8.5cm]{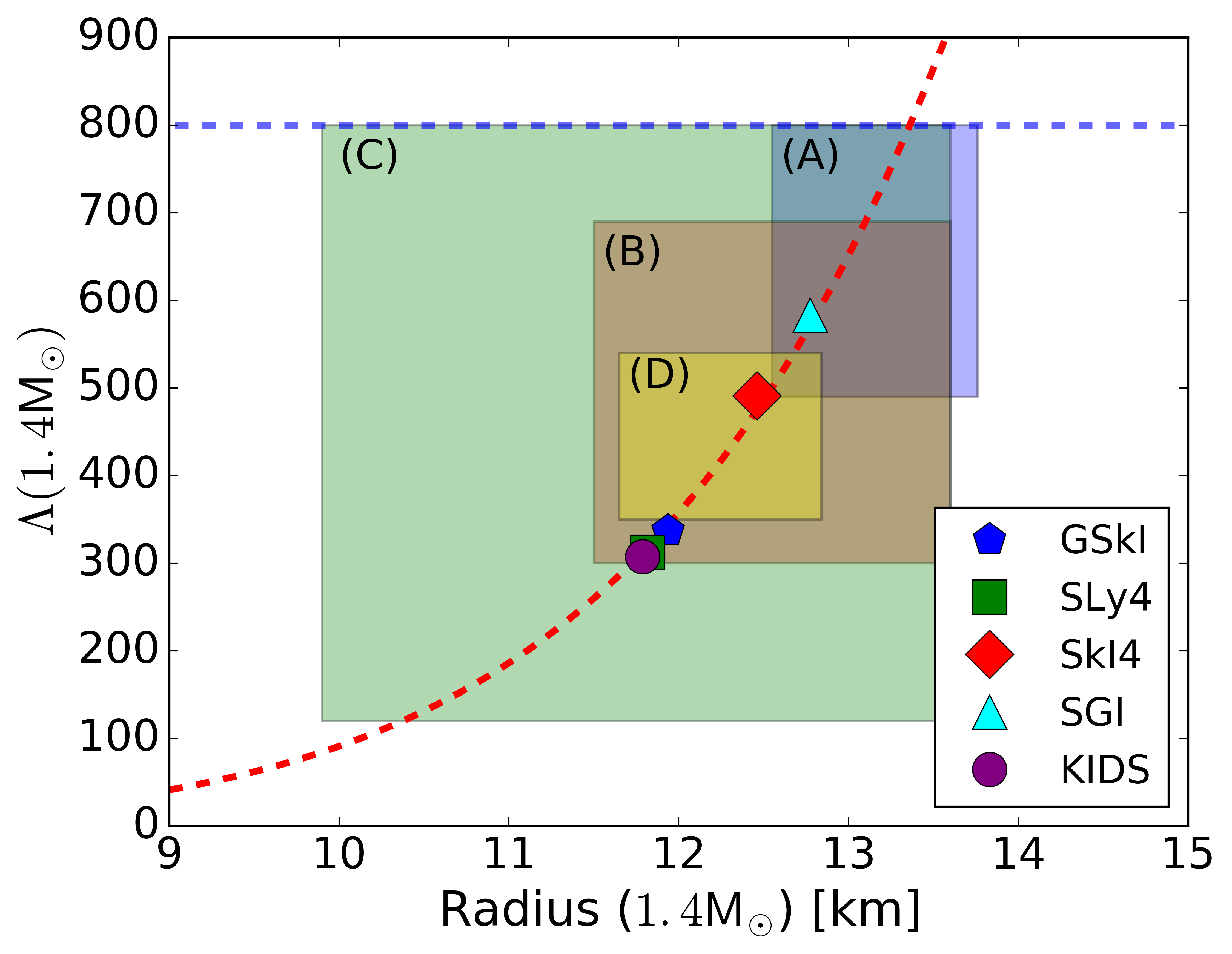}
\caption{Dimensionless tidal deformability ($\Lambda$) of a single neutron star as a function of radius when the mass of neutron star is fixed at $1.4~M_{\odot}$. The results from our five models are indicated with color symbols. 
The shaded rectangles are adopted from the following references: purple (A) \cite{fattoyev}, brown (B) \cite{krastev}, green (C) \cite{annala}, and gold (D) \cite{lim}. 
%They represent a 90$\sim$95\% confidence interval for each R and $\Lambda$ independently, i.e., without any correlation between R and $\Lambda$. 
The red dashed curve corresponds to $\Lambda \sim R^{7.5}$ shown in \cite{annala} and it overlaps the results from our five models very well. The horizontal blue dashed line ($\Lambda=800$) indicates again the upper limit of $\Lambda$ estimated from GW170817 \cite{170817}.} 
\label{fig3}
\end{figure}

Since the first measurement of tidal deformability ($\Lambda$ and $\tilde{\Lambda}$) from GW170817, 
several papers have been 
published already comparing predictions from various models with the measured values. Because 
it is worth comparing our results obtained from the five selected EoS models with those from previous works, we review some of recent works~\cite{annala,fattoyev,krastev,lim} and compare our results with them in the following.

By using tri-tropic and quadru-tropic functions, Ref.~\cite{annala} constructs more than 20,000 EoSs
which are interpolated between the low-density EoS computed with chiral effective field theory \cite{eft} and
the high-density EoS obtained from non-perturbative QCD \cite{pqcd}. These EoSs are tested 
against the upper limit of $\Lambda(1.4M_\odot) < 800$ and the maximum mass of a neutron star
($M_{\rm max} > 2.0 M_\odot$). 
Radius of a neutron star with $1.4 M_\odot$ ($R_{1.4M_\odot}$) is found to be 
bound between 11.1~km and 13.4~km, consistent with the constraint of the maximum mass and the tidal deformability. 
If the upper limit of $\Lambda(1.4M_\odot)$ were lowered to 400, 
the radius of the $1.4 M_\odot$ neutron star could decrease to 12.3~km. 
The results obtained for the selected models in our work are in good agreement with those in \cite{annala}.

In Ref.~\cite{fattoyev}, the EoS models based upon the relativistic mean field (RMF) theory are considered. Ten models 
are selected first, all of which satisfy the maximum mass of neutron star ($\sim 2 M_\odot$), and 
tested against the tidal deformability. For the additional test, they also include measurement of 
the neutron skin thickness of $^{208}$Pb ($R^{208}_{np}$).
This measurement was done by the Lead Radius 
Experiment (PREX) at the Jefferson Laboratory \cite{prex1,prex2}. 
Because the radius of a neutron star with $1.4 M_\odot$ ($R_{1.4M_\odot}$), 
tidal deformability ($\Lambda(1.4M_\odot)$ and $\tilde{\Lambda}$), and 
$R^{208}_{np}$ are predicted together by a given EoS, 
their correlation can be used to constrain each other.  
When the upper limit of $\Lambda(1.4M_\odot) < 800$ is applied to ten models which predict 
various radii at a given neutron star mass of $1.4M_\odot$, only four models having relatively soft EoS pass the test. 
By applying $\Lambda(1.4M_\odot) < 800$ to a radius function fitted to ten models, 
$R_{1.4M_\odot}$ is estimated less than $13.76$~km.  
Note that this limit on $R_{1.4M_\odot}$ is in good agreement with ours and that in Ref.~\cite{annala}.
The condition $\Lambda(1.4M_\odot) < 800$ also sets an upper limit on the neutron skin thickness
as $R^{208}_{np} < (0.25 \sim 0.28)$~fm, which is obtained with predictions from ten models. 
However, PREX reports a wide range of measured values, $R^{208}_{np}= 0.33^{+0.16}_{-0.18}$~fm
\cite{prex1,prex2}. It is interesting that a substantial portion of PREX value is ruled out 
with the constraint of $\Lambda(1.4M_\odot) < 800$. 
Alternatively, the lower limit of PREX data ($R^{208}_{np} \approx 0.15$~fm) can provide 
a constraint on the minimum value of $\Lambda(1.4M_\odot)$. % $\tilde{\Lambda}$.
Fitted to predictions from ten models, the lower limit on $R^{208}_{np}$ leads to 
$R_{1.4M_\odot} \approx 12.55$~km and 
$\Lambda(1.4M_\odot) \approx 490$. If an equal mass binary were assumed, 
$\tilde{\Lambda} = \Lambda(1.4M_\odot) \approx 490$. 
This lower limit on $\tilde{\Lambda}$ is larger than 400 which is estimated from AT2017gfo \cite{at2017}. 
Note that the lower limit calculated from the models in this work is about 300 
as shown in Table~\ref{table:radius}.

Ref.~\cite{krastev} focuses on the role of symmetry energy in EoS.
Employing the momentum-dependent interaction (MDI) model, 
and varying the parameter $x$ in the model which 
controls stiffness of the EoS such as 
$x=0$ (softest limit) to $x=-1$ (stiffest limit), $300 \leq \tilde{\Lambda} \leq 690$ is obtained 
for an equal mass binary composed of two $1.4 M_\odot$ stars. Note that the lower (upper) limit 
on $\tilde{\Lambda}$ comes from $x=0$ ($x=-1$). 
When applied to a single neutron star, 
the MDI model predicts $11.5~{\rm km} \leq R_{1.4M_\odot} \leq 13.6~{\rm km}$ which also corresponds to between $x=0$ (lower 
limit) and $x=-1$ (upper limit).  
Note that the range of the tidal deformability predicted in \cite{krastev} is consistent 
with the results from GW170817 and AT2017gfo, 
and that the range of $R_{1.4M_\odot}$ is similar to that obtained in \cite{annala}.

Ref.~\cite{lim} calculates the tidal deformability with the EoS constrained by chiral 
effective field theory up to twice nuclear saturation denisty.
The effect of uncertainties from high orders and dependence on the momentum cutoff 
values are analyzed statistically.
For the $1.4 M_\odot$ mass neutron star, they obtain the radius in the range
$11.65 \leq R_{1.4 M_\odot} \leq 12.84$~km, and the single star tidal deformability
$375.51 \leq \Lambda (1.4 M_\odot) \leq 538.43$ at the $2 \sigma$ (95\%) confidence
level.

Fig.~\ref{fig3} shows the tidal deformability $\Lambda(1.4 M_\odot)$ from the models
discussed above. 
The results from our five selected models are in good agreement with those from the other models~\cite{annala,fattoyev,krastev,lim}. In particular, all the data points of the five selected models 
are on the curve ($\Lambda \sim R^{7.5}$) from \cite{annala}. 
Our results are also consistent with the lower bounds on the deformability and radii of neutron stars estimated using the gravitational waves based on the Bayesian analysis~\cite{Deetal}.

\section{Summary}

The first detection of gravitation wave from the merge of neutron star binary (GW170817) provided a valuable information on 
tidal deformability of merging neutron stars which is determined by the EoS of dense matter inside the neutron star.   
In order to investigate whether the measured tidal deformability can be predicted with reasonable EoS models which 
have been tested already against the other sets of criteria, we select five specific EoS models  
based upon the Skyrme force model and density functional theory. The selected models are SkI4, SGI, SLy4, 
GSkI, and KIDS and their predictions are in good agreement with three criteria, data of finite nuclei, nuclear matter, and 
the maximum mass of neutron star. For these five models, dimensionless tidal deformability ($\Lambda$) of a single
neutron star and mass--weighted tidal deformability ($\tilde{\Lambda}$) of the binary neutron stars are calculated. 
For a single neutron star with 
$1.4 M_\odot$, our five models predict a range of radius $11.8~{\rm km} \leq R_{1.4M_\odot} \leq 12.8~{\rm km}$ and 
a range of dimensionless tidal deformability $308 \leq \Lambda(1.4 M_\odot) \leq 583$ which satisfies the 
measured upper limit  $\Lambda(1.4 M_\odot) \leq800$ (see Fig.~\ref{fig2}, \ref{fig3} and Table~\ref{table:radius}). 
For the mass-weighted tidal deformability ($\tilde{\Lambda}$) 
of the binary neutron stars, our five models predict $360 \leq \tilde{\Lambda} \leq 700$ for the fixed chirp mass of 
 $M_{\rm chirp}=1.188 M_{\odot}$ and a range of mass ratio $0.7 \leq q = M_2/M_1 \leq 1.0$ which are also 
 estimated from the GW170817 observations for the low-spin prior (see Fig.~\ref{fig4}). 
 The predicted range of $\tilde{\Lambda}$ also 
 satisfies the measured upper limit $\tilde{\Lambda} \leq 800$, but the lower value of the predicted range is 
 marginal in comparison with the estimated lower limit $\tilde{\Lambda} \geq 400$ from AT2017gfo in \cite{at2017}. 
 We compare our results with those of recent works and find that our results are in good agreement with them~\cite{annala,fattoyev,krastev,lim,Deetal} while 
 providing a narrower range of predictions due to a smaller number of models in consideration. 
 From our results, however, we can conclude that the first measured tidal deformability from GW170817 
can be also explained by the tested EoSs which have been successfully explaining the 
properties of finite nuclei and nuclear matter as well as the maximum mass of neutron star.   

\section*{Acknowledgments}

YMK and CHL were partially supported by the National Research Foundation of Korea (NRF) grant funded by the 
Korea government (MSIP) (No. 2015R1A2A2A01004238 and No. 2016R1A5A1013277). 
YL was supported by National Science Foundation under Grant No. PHY1652199. 
KK was supported by NRF grant by MSIP (No. 2016R1A5A1013277). CHH was supported by the Basic Science Research Program through the NRF funded by the Ministry of Education (NRF-2017R1D1A1B03029020).


\begin{thebibliography}{99}

\bibitem{170817}
B.~P. Abbott {\it et al.},
Phys. Rev. Lett. {\bf 119}, 161101 (2017).
%
\bibitem{170817GRB}
A. Goldstein {\it et al.},
Astrophys. J. Lett.  {\bf 848}, 14 (2017).
%
\bibitem{170817AT}
B.~P. Abbott {\it et al.},
Astrophys. J. Lett.  {\bf 848}, 12 (2017).
%
\bibitem{flanagan2008}
E.~E.~ Flanagan, and T. Hinderer,
Phys. Rev. D {\bf 77}, 021502 (2008).
%
\bibitem{hinderer2008}
T. Hinderer,
Astrophys. J. {\bf 677}, 1216 (2008).
%
\bibitem{a}
P. Demorest, T. Pennucci, R. Ransom, M. Roberts, and J.~W.~T. Hessels,
Nature {\bf 467}, 1081 (2010).
%
\bibitem{b}
J. Antoniadis {\it et al.},
Science {\bf 340}, 448 (2013).
%
\bibitem{at2017}
D. Radice, A. Perego, F. Zappa, and S. Bernuzzi,
arXiv:1711.03647v3 [astro-ph.HE].
%
\bibitem{dutra}
M.~Dutra, O.~Lourenco, J.~S.~S\'{a} Martins, and A.~Delfino,
%\newblock Skyrme Interactions and Nuclear Matter Constraints,
\newblock Phys. Rev. C \textbf{85}, 035201 (2012).
%
\bibitem{ski4}
P.-G. Reinhard, and H. Flocard,
Nucl. Phys. {\bf A584}, 467 (1995).
%
\bibitem{sg1}
Guo-Qiang Li,
J. Phys. G {\bf 17}, 1 (1991).
%
\bibitem{sly4}
E. Chabanat, P. Bonche, P. Haensel, J. Meyer, and R. Shaeffer,
Nucle. Phys. {\bf A635}, 231 (1998).
%
\bibitem{gsk1}
B.~K. Agrawal, S.~K. Dhiman, and R. Kumar,
Phys. Rev. C {\bf 73}, 034319 (2006).

\bibitem{prc2018}
P. Papakonstantinou, T.-S. Park, Y. Lim, and C.~H. Hyun,
%Density dependence of the nuclear energy-density functional,
Phys. Rev. C {\bf 97}, 014312 (2018).

\bibitem{Hin10}
T. Hinderer, B. D. Lackey, R. N. Lang, and J. S. Read, Phys. Rev. D {\bf 81}, 123016 (2010).

\bibitem{Fav14}
M. Favata, Phys. Rev. Lett. {\bf 112}, 101101 (2014).
%
%\bibitem{ijmpe2015}
%Y. Lim, C.~H. Hyun, K. Kwak, and C.-H. Lee,
%Int. J. Mod. Phys. E {\bf 24}, 1550100 (2015).
%

%
\bibitem{acta2017}
H. Gil, P. Papakonstantinou, C.~H. Hyun, T.-S. Park, and Y. Oh,
%Nuclear energy density functional for KIDS,
Acta. Phys. Polon. B {\bf 48}, 305 (2017).
%
\bibitem{sm2017}
H. Gil, P. Papakonstantinou, C.~H. Hyun, and Y. Oh,
%Skyrme-type nuclear force for the KIDS energy density functional,
New Physics: Sae Mulli {\bf 67}, 456 (2017).
%
\bibitem{lns2006}
L. G. Cao, U. Lombardo, C. W. Shen, and N. V. Giai, 
Phys. Rev C {\bf 73}, 014313 (2006).
%
\bibitem{steiner2010}
A.~W. Steiner, J.~M. Lattimer, and E.~F. Brown,
Astrophys. J. 722, 33 (2010).
%
\bibitem{annala}
E. Annala, T. Gorda, A. Kurkela, and A. Vuorinen,
arXiv:1711.02644v1 [astro-ph.HE].
%
\bibitem{fattoyev}
F.~J. Fattoyev, J. Piekarewicz, and C.~J. Horowitz,
arXiv:1711.06615v2 [nucl-th].
%
\bibitem{krastev}
P.~G. Krastev, and B.-A. Li,
arXiv:1801.04620v1 [nucl-th].
%
\bibitem{lim}
Y. Lim amd J. Holt,
arXiv:1803.02803 [nucl-th].
%
\bibitem{eft}
I. Tews, T. Krger, K. Hebeler, and A. Schwenk,
Phys. Rev. Lett. {\bf 110}, 032504 (2013).
%
\bibitem{pqcd}
A. Kurkela, P. Romatschke, and A. Vuorinen,
Phys. Rev. D {\bf 81}, 105021 (2010).

%
\bibitem{prex1}
S. Abrahamyan, {\it et al.},
Phys. Rev. Lett. {\bf 108}, 112502 (2012).
%
\bibitem{prex2}
C.~J. Horowitz {\it et al.},
Phys. Rev. C {\bf 85}, 032501 (2012).

%\bibitem{demorest2010}
%P. Demorest, {\it et al.},
%Nature 467, 1081 (2010)

%\bibitem{antoniadis2013}
%J. Anoniadis, {\it et al.},
%Science 340, 1233232 (2013)

\bibitem{Deetal}
S. De, D. Finstad, J.M. Lattimer, {\it et al.}, arXiv:1804.08583 (2018).

 
\end{thebibliography}
\end{document}